\documentclass[twocolumn]{aa}
\usepackage{graphicx, subfigure}
\usepackage{txfonts}
\begin{document}
   \title{3D spectroscopy with VLT/GIRAFFE - II: Are Luminous Compact
   Galaxies merger remnants ?\thanks{Based on observations collected
   at the European Southern Observatory, Paranal, Chile, ESO Nos
   71.A-0322(A) and 72.A-0169(A)}}

   %\subtitle{FLAMES/GIRAFFE Paris Observatory GTO\thanks{Guaranteed Time Observations} - II}

   \author{M. Puech, F. Hammer, H. Flores\inst{1}, G. \"Ostlin\inst{2}
   \and T. Marquart\inst{3}}

   \titlerunning{Are LCGs merger remnants ?}

   \authorrunning{M. Puech, F. Hammer, H. Flores et al.}

   \offprints{mathieu.puech@obspm.fr}

   \institute{Laboratoire Galaxies Etoiles Physique et
        Instrumentation, Observatoire de Paris, 5 place Jules Janssen,
        92195 Meudon France \and Stockholm Observatory, AlbaNova
        University Center, 10691 Stockholm, Sweden \and Dept. of
        Astronomy and Space Physics, Box 515, 75120 Uppsala, Sweden}

   \date{Received xxx, 2005; accepted xxx, 2005}

   \abstract{

Luminous Compact Galaxies are enigmatic sources by many aspects. They
can reach the luminosity of the Milky Way within a radius of only a
few kpc. They also represent one of the most rapidly evolving
populations of galaxies since they represent up to 1/5 of the luminous
galaxies at redshift $z= 0.7$ while being almost absent in the local
Universe. The measurement of their dynamics is crucial to our
understanding of LCGs since this has the potential of telling us which
physical process(es) that drives them, and ultimately to link them to
the existing present-day galaxies. Here we derive the 3 dimensional
velocity fields and velocity dispersion ($\sigma$) maps of 17 Luminous
Compact Galaxies selected from the Canada France Redshift Survey and
the Hubble Deep Field South with redshifts ranging from z=0.4 to
z=0.75. We find that only 18\% of them show rotational velocity fields
typical of rotating disks, the others showing more complex kinematics.
Assuming that LCGs are not too far from equilibrium, about half of
LCGs then appear to be either non-relaxed objects, or objects that are
not supported by velocity dispersion alone. This supports the view
that an important fraction of LCGs are probably mergers. It brings
additional support to the ``spiral rebuilding scenario'' in which LCGs
correspond to a previous or post-merger phase before the disk
re-building.

   }

   \maketitle

%________________________________________________________________
\section{Introduction}
Luminous Compact Galaxies (LCGs) (\(M_B<-20\), \(r_{half}<5\)
\(h_{50}^{-1}\) kpc and $EW_0([OII]) >$15\AA) correspond to the most
rapidly evolving population seen in the UV (Lilly et al. 1998): they
represent $\sim$ 20\% of the \(1>z>0.4\) galaxies (Zheng et al. 2005),
$\sim$50\% of the emission line galaxies (see Paper I, Flores et al.
2006) and almost vanish in the local Universe, with their number
density decreasing by factors 7--10 (Jangren et al. 2004; Garland et
al. 2003; Werk et al. 2004). Moreover, LCGs contribute to 40--50\% of
the increase in the cosmic Star Formation Rate (SFR) density between
z=0 and 1 as measured from rest frame UV luminosities (Lilly et al.
1995; Guzman et al. 1997) and 25\% in the IR (Zheng et al. 2004). The
spectra of LCGs reveal a mixture of old, intermediate and young
stellar populations (Hammer et al. 2001). Apart from their
compactness, LCGs have properties surprisingly similar to those of
other, more extended luminous IR galaxies and starbursts: They show a
similar mix of stellar populations, extinction property distributions,
stellar masses and SFRs (Hammer et al. 2005). This led Hammer et al.
(2001) to propose that LCGs are the progenitors of present-day bulges
of early type spirals.
%However, since most of the E/S0 population
%seems to be well in place by z=1, it appears unlikely that many
%present-day ellipticals have formed recently by this process.

Recently, Hammer et al. (2005) proposed the so-called ``spiral
rebuilding scenario'' to explain the formation of the 75\% most
massive local spirals. This scenario is composed of 3 major phases: a
``pre-merger phase'' during which two distant spirals merge, the ``LCG
phase'' where all material from the progenitors fall into the mass
barycenter of the system and form a bulge, and the ``disk growing
phase'' where subsequently accreted material form a rotating disk.
This scenario is partly supported by \"Ostlin et al. (1999 and 2001)
who obtained very complex H\(\alpha\) velocity fields for local LCGs,
characteristic of what is expected from merging galaxies. Another
alternative is the one proposed by Barton and Van Zee (2001):
comparing HI and optical emission line widths of nearby LCGs
candidates they argued that interactions and minor mergers of disk
galaxies may cause apparently compact morphology leading too to the
formation of a bulge. Both views are not incompatible and can occur
during different stages of the ``LCG phase'' described by Hammer et
al. (2005).

Compact galaxies have been extensively studied in the past. Koo et al.
(1995) were the first to propose that some compact galaxies could be
the progenitors of local dwarf ellipticals (dE), assuming that they
experience a dramatic event of star formation before fading away by up
to 5 magnitudes. Guzman et al. (1997) established a distinction
between two types of compact galaxies: 60\% present properties
characteristic of young star-forming HII galaxies (in e.g. velocity
widths, SFRs and mass-to-light ratios), whereas the remaining 40\%
constitute a more heterogeneous class of evolved starbursts, similar
to local starburst disk galaxies. Philips et al. (1997) then suggested
that the HII-like compact galaxies are the best candidates to evolve
into dEs. In this paper we study the most Luminous and most actively
star forming fraction of the population of Compact Galaxies (LCGs),
i.e. those which contribute most to the increase in the star formation
rate density. It is important here to stress that the sample presented
in this paper correspond to the brightest 25\% of galaxies in the
sample of Blue Compact Galaxies (BCGs) in the Hubble Deep Field (HDF)
at \(0.1 < z <1.3\) studied by Guzman et al. (1997)\footnote{They
defined a compact galaxy as \(r_{half}\leq 0.5\) arcsec and
\(SB_{I814}\leq 22.2\) mag.arcsec\(^{-2}\).}, and would have mostly
been classified as SB-disk like compact galaxies following their
criteria. They thus do not correspond to the sub-class of compact
galaxies for which Guzman et al. (1997) would have considered as
possible progenitors of dEs. Notice that for another sub-class of
compact galaxies, Philips et al. (1997) found that ``\emph{one
possibility is that they are disks forming from the center outward,
and so the radius of the luminous material and enclosed mass are small
compared to present-day spirals}''.

%In this paper we study the most massive and most actively star forming
%fraction of the population of Compact Galaxies, i.e. those which
%contribute most to the increase in the star formation rate density.
%Our sample correspond to the brightest 25\% of galaxies in the sample
%of Blue Compact Galaxies (BCGs) in the Hubble Deep Field (HDF) at
%\(0.1 < z <1.3\) studied by Guzman et al. (1997)\footnote{They defined
%a compact galaxy as (\(r_{half}\leq 0.5\) arcsec and \(SB_{I814}\leq
%22.2\) mag.arcsec\(^{-2}\))}.

%Koo et al. (1995, see also Guzman et al. 1997; Bershady et al. 2004)
%suggested that distant compact galaxies could be the progenitors of local
%dwarf ellipticals (like NGC 205), assumed to experience a dramatic
%event of star formation before fading away by up to 5 magnitudes. A
%major motivation for this scenario was the small velocity widths 
%of compact galaxies, a property also shared by several of the
%luminous ones studied by Hammer et al (2001). The "dwarf elliptical
%progenitor" picture is however challenged by the prominent stellar
%population of LCGs with old and intermediate ages and by their stellar
%and metal content which is far greater than that of dwarf ellipticals.

Whereas the LCGs are important for understanding galaxy evolution
since $z=1$, their nature is thus still enigmatic. The aim of this
paper is to evaluate the nature of their kinematics from a survey of
17 LCGs randomly selected from the CFRS and the HDF-South (HDFS) field
at redshifts from $z=$0.4 to 0.75, and to investigate if their
dynamics is supported either by rotation or velocity dispersion. This
will help in distinguishing a merger scenario from a dwarf elliptical
experiencing a 5 magnitude brightening. Section 2 present the sample,
observations and the methodology we followed. Kinematical and
dynamical results are in section 3 and 4. We discuss our results in
section 5 and a conclusion is given in section 6. In the following, we
assumed a $\Lambda$-CDM cosmology with $H_0=70$, $\Omega_m=0.3$,
$\Lambda_0=0.7$ and $q_0=-0.55$.

%__________________________________________________________________
\section{Sample, observations \& methodology}

To select compact galaxies, we used Hubble Space Telescope (HST)
images in the F814W filter (WFPC2, 0.1 arcsec/pix; and ACS, 0.05
arcsec/pix) in the CFRS (3hr and 22hr fields) and the HDFS field. For
one galaxy, however, we used ground-based images obtained by the CFRS
team at the CFHT (0.207 arcsec/pix, see Hammer et al. 2001).

We selected 21 LCGs with \(0.4\leq z\leq 0.75\), following the
procedure detailed in Hammer et al. (2001), using the light
concentration parameter $\delta$ as a compactness criterion which
measures the difference between the luminosities within the 5 and 15
kpc radii (see Table \ref{tab1}). The condition $\delta I < 0.73$
allows to select galaxies with $r_{half} \leq 5 h^{-1}_{50}$ kpc in a
homogeneous way. This value corresponds to $\sim 4.34$ kpc in a
$\Lambda$-CDM cosmology at $z\sim0.55$. To achieve a homogeneous
selection, ACS images were degraded and re-sampled to the WFPC2
quality using a point spread function (PSF) generated with the Tiny
Tim software. This size criterion ensures the selection of relatively
small galaxies (see Ravindranath et al. 2004), although not
necessarily as compact as those selected by some other authors (e.g.
Guzman et al. 1997). Half light radii were then derived by
interpolating luminosities enclosed within concentric ellipses using
the IRAF polyphot task (see Hammer et al. 2001 for a complete
description of the procedure). Inclinations were estimated using
Sextractor (Bertin et al. 1996) and the ellipse task of IRAF: we find
a mean absolute difference of $\sim 2$ degrees between these two
methods. Independent measurements were also done by eye and gave
similar results as Sextractor to within $\sim 4$ degrees. In the
following, we will use the estimates from Sextractor and assume an
error of $\pm 4$ degrees.

As part of the Guaranteed Time Observations (GTO) of the Paris
Observatory, we obtained observations with the FLAMES/GIRAFFE
instrument of the 21 compact galaxies using the deployable integral
field units (IFUs), each covering an area of 3 by 2 arcsec$^2$, at
0.52 arcsec/pixel. The complete description of the GTO sample is given
in Paper I (Flores et al. 2006). Briefly, we used LR04 and LR05 setups
targeting the [OII] doublet (R\(\sim10000\)), integration times
ranging from 8 to 13 hours and the seeing was typically \(\sim 0.6\)
arcsec during all the observations. Data cubes were reduced using the
GIRBLDRS v1.12 package (Blecha et al. 2000), including narrow
flat-fielding. Sky was carefully subtracted with our own IDL
procedures.

Among these 21 compact galaxies, we selected 17 galaxies for which at
least 4 pixels had [OII] doublet reaching a spectral signal to noise
ratio (SNR) of 4 (see definition in Paper I). In the following we
focus only on these 17 remaining LCGs (see Table \ref{tab1}). At first
sight, it might seem too challenging to derive velocity fields of
compact galaxies ($r_{half}$ $\sim$ 0.5 arcsec), using the GIRAFFE/IFU
with 0.52 arcsec microlenses. To assess this we examine how many
pixels that reach an integrated SNR of 3. Among the sample of LCGs, we
found a median value of 11 pixels ($\sim$ 3 arcsec$^2$) compared to 16
($\sim$ 4.3 arcsec$^2$) for a sample of 8 spirals of Paper I (Flores
et al. 2006). On average our LCGs are thus $\sim$ 30\% less extended
than spirals. However, the average filling factor of the IFU (20
pixels) for the sample of 17 LCGs is $\sim$ 55 \%, which is sufficient
to explore the kinematics of these galaxies.

We processed the spectra with a Savitzky-Golay filtering, which has
the advantage over the widely used box smoothing that it conserves the
first moment of spectral lines (Press et al. 1989). We identified the
[OII] doublet by visual inspection and retained the spectra that
reached a sufficient spectral SNR of 3. We then fitted a double
Gaussian with the following constraints (where the subscripts denote
the two components of the fit): $\lambda _2$-$\lambda _1$=2.783$\AA$
(in rest frame wavelength) and $\sigma_1$=$\sigma _2$. The line ratio
was allowed to vary freely except when the fit failed: in these cases
we forced the line ratio to a value of 1.4 which was the median value
observed in our integrated spectra. This occurs for pixels with rather
low SNR and affects $\sim$ 10 \% of the measured pixels and will thus
not significantly affect our results. In all such cases, we checked by
eye that the derived fit was acceptable and took this into account
during the classification (see below). The complete procedure is
described in Paper I.

Figure \ref{Fig1} shows velocity and velocity dispersion ($\sigma$)
maps for the 17 LCGs. The $\sigma$ -maps were corrected for the
instrumental broadening using sky lines. To make the interpretation
easier, velocity fields and $\sigma$-maps are presented after a simple
5x5 linear interpolation. 
%We also show SNR maps where the pixels are
%divided in 4 classes: very high (integrated SNR $\geq$ 38), high (25
%$\leq$ integrated SNR $\leq$ 38), low (19 $\leq$ integrated SNR $\leq$
%25) and bad (integrated SNR $\leq$ 19, discarded). 
%For each galaxy we also show a SNR map and compute a quality factor Q
%by adding the number of pixels in four SNR classes (very high with an
%integrated SNR $\geq$ 38, high with 25 $\leq$ integrated SNR $\leq$
%38, low with 19 $\leq$ integrated SNR $\leq$ 25 and bad with an
%integrated SNR $\leq$ 19, which have been discarded) multiplied by the
%corresponding thresholds, and divided the result by the number of
%spectral resolution elements in the doublet (see Table \ref{tab1}).

{\scriptsize
\begin{table*}
\centering
\begin{tabular}{lcccccccrr}\hline\hline
ID    & z & I$_{AB}^a$ &  M$_B^b$  &  M$_K^b$ & $\delta I$ & r$_{half}^c$ & i$^d$\\\hline
HDFS4170       & 0.4602  & 20.79  & -20.43 &  -22.60 & 0.56  & 3.57 &  51\\
HDFS5190       & 0.6952  & 21.31  & -21.25 &  -21.92 & 0.60  & 4.07 &  59\\
CFRS03.0619    & 0.4854  & 20.80  & -20.67 &  -21.93 & 0.63  & 3.87 &  27\\
CFRS03.1032    & 0.6180  & 20.49  & -21.18 &  -22.64 & 0.51  & 1.79 &  37\\
CFRS22.0619    & 0.4676  & 21.55  & -19.33 &  -19.33 & 0.70  & 4.31 &  68\\
CFRS03.1349    & 0.6155  & 20.87  & -21.19 &  -22.92 & 0.56  & 3.84 &  48\\
CFRS22.1064    & 0.5383  & 22.08  & -19.87 &  -21.64 & 0.24  & 2.36 &  49\\
HDFS5150       & 0.6956  & 22.36  & -20.20 &  -21.03 & 0.72  & 3.39 &  42\\
CFRS03.0508    & 0.4642  & 21.92  & -19.61 &  -20.35 & 0.46  & 3.32 &  38\\
CFRS03.0645    & 0.5275  & 21.36  & -20.30 &  -21.34 & 0.71  & 4.57 &  45\\
CFRS22.0919    & 0.4738  & 21.77  & -19.99 &  -19.53 & 0.41  & 2.52 &  61\\
CFRS22.0975    & 0.4211  & 20.21  & -20.40 &  -22.53 & 0.59  & 3.82 &  50\\
CFRS03.0523    & 0.6508  & 21.31  & -20.67 &  -21.55 & 0.48  & 3.57 &  41\\
HDFS4130       & 0.4054  & 20.09  & -20.90 &  -22.13 & 0.62  & 4.04 &  36\\
HDFS4090       & 0.5162  & 22.15  & -19.71 &  -19.83 & 0.35  & 1.54 &  45\\
HDFS5140       & 0.5649  & 22.38  & -19.76 &  -20.46 & 0.36  & 2.57 &  50\\
HDFS5030       & 0.5821  & 20.40  & -21.74 &  -22.68 & 0.66  & 4.19 &  25\\\hline\hline
\end{tabular}
\begin{list}{}{}
\item[$^{\mathrm{a}}$] isophotal magnitudes.
\item[$^{\mathrm{b}}$] from Hammer et al. 2005.
\item[$^{\mathrm{c}}$] in kpc.
\item[$^{\mathrm{d}}$] inclination, in deg.
%\item[$^{\mathrm{e}}$] number of pixels in the IFU with a SNR lower than 19.
\end{list}
\caption{Main properties of the sample of LCGs: galaxies names,
redshifts, isophotal I magnitudes, absolute B and K magnitudes, light
concentration, half light radii, inclination.}
\label{tab1}
\end{table*}
}

%__________________________________________________________________
\section{Kinematics of LCGs}
Following Paper I (Flores et al. 2006), we define 3 kinematical
 classes: rotating disks (velocity field showing rotation and
 $\sigma$-map showing a peak near the center), perturbed rotations
 (velocity field showing rotation but $\sigma$-map without peak or
 with a peak offset from the center) and complex kinematics, (see
 Figure \ref{Fig1} and comments on individual objects). This
 classification relies on the fact that observations with a low
 spatial resolution integral field spectrograph, such as GIRAFFE,
 should reveal line widths dominated by the integration of larger
 scale motions and not by intrinsic random motions: in the case of a
 rotating disk, $\sigma$ should show a peak near the galaxy center,
 where the gradient of the rotation curve is the steepest (e.g. Van
 Zee \& Bryant 1999). This classification has been checked through
 numerical simulations which are described in detail in Paper I. In
 these simulations, we assumed that \emph{all} the observed galaxies
 are indeed rotating disks (taken as a standard model) and that all
 the observed large scale motions in the velocity fields correspond to
 rotations. In other words, we have tried to force each system to
 appear as rotational disks and then estimate the discrepancy between
 the observed system and the adopted standard rotation (see Paper I).
 Below we comment on the velocity fields of individual objects.

\begin{figure*}[h!]
\centering
\subfigure{
}
\caption{SEE ATTACHED JPEG FIGURE Kinematics of 17 LCGs. From left to right: I band HST imaging
  (0.1 arcsec/pix, FoV=3x2 arcsec$^2$) with the IFU bundle
  superimposed; velocity field (5x5 interpolation, $\sim 0.1$
  arcsec/pix), $\sigma$-map (5x5 interpolation, $\sim 0.1$ arcsec/pix)
  and SNR map (mean spectral SNR, see Paper I). From up to down: LCGs
  classified as rotating disks (3 first), perturbed (5 next) and
  complex (9 last). The 3 first galaxies classified as complex
  (CFRS03.0508, CFRS03.0645 and HDFS5140) are those suspected to be
  dominated by outflows.}
\end{figure*}

\addtocounter{figure}{-1}
\begin{figure*}[h!]
\centering
\subfigure{
}
\caption{SEE ATTACHED JPEG FIGURE \emph{continued.}}
\label{Fig1}
\end{figure*}

\begin{description}
\item{{\bf HDFS4170}}

Its kinematics is classified as a rotating disk: a rotation is seen in the
velocity field and the $\sigma$-map has a clear peak in the center.\\

\item{{\bf HDFS5190}}

Its kinematics is classified as a rotating disk: a clear rotation is
seen in the velocity field and the $\sigma$-map shows a peak near the
center.\\

\item{{\bf CFRS03.0619}}

Its kinematics is classified as a rotating disk: the velocity field
shows a clear rotation and the $\sigma$-map has a peak slightly offset
from the center of the galaxy. Due to the low spatial resolution of
GIRAFFE, we choose to classify this galaxy as rotating disk rather
than perturbed rotation.\\

\item{{\bf CFRS03.1032}}

Its kinematics was first classified as complex because the velocity
field looked perturbed and the $\sigma$-map has a peak near the edge
of the galaxy. Although our simulation cannot reproduce the amplitude
of the $\sigma$ peak, this peak is however located in the same pixel
as the one seen in the observed $\sigma$-map. HST/ACS imaging reveals
a very compact structure, completely dominated by the center. This
galaxy is the most compact in the sample. We retrieved a FORS slit
spectrum from which we found a Log([OIII]/H$_\beta$) ratio of 1.8,
which characterize a Seyfert galaxy spectrum: the high $\sigma$ value
of this galaxy might reflects an AGN activity. We choose to
re-classify its kinematics as a perturbed rotation, but we notice that
it somewhat escapes our classification scheme. \\

\item{{\bf CFRS22.0619}}

Its kinematics is classified as a perturbed rotation: a clear rotation
is seen in the velocity field but the $\sigma$-map has a peak at the
edge of the galaxy. Note that this galaxy is seen nearly edge-on.\\

\item{{\bf CFRS03.1349}}

Its kinematics was first classified as a rotating disk because a clear
rotation is seen in the velocity field and a peak is seen near the
center of the $\sigma$-map. However, the simulation (see Paper I)
cannot reproduce the location of the $\sigma$ peak, although the
secondary peak (at the bottom-left of the maximal peak) is reproduced.
Zheng et al. (2004) classified this galaxy as Sab with a compact
bulge, relatively blue compared to the disk. An interacting companion
is 20 kpc away at the same redshift. Interestingly, this galaxy shows
a companion galaxy at $\sim$7 kpc, and the distorsion in the $\sigma$
map is oriented towards this companion. We suspect that an interaction
(gas falling) is responsible for both the star formation activity and
distorsion of the kinematics (following the scenario proposed by
Barton \& Van Zee 2001). We then choose to re-classify its kinematics
as a perturbed rotation.\\

\item{{\bf CFRS22.1064}}

Its kinematics is classified as perturbed: the velocity field shows 
rotation but the $\sigma$-map is very perturbed.\\

\item{{\bf HDFS5150}}

Its kinematics is classified as a perturbed rotation: rotation is seen
in the velocity field but the $\sigma$-map has a peak at the edge of
the galaxy. The morphology looks quite irregular.\\

\item{\bf {CFRS03.0508}}

The velocity field  clearly shows an apparent rotation and the
$\sigma$-map has a well-defined peak in the center. Note however that
the dynamical axis seems almost orthogonal to the photometric axis of
the brightest component: this could be a signature of outflows
(Veilleux et al. 2005; Bosma, private communication). Hence its
kinematics has been classified complex. From morphological studies,
Zheng et al. (2005) classified this galaxy as the relics of an interaction 
or merger, with a relatively blue color over the whole galaxy.\\

\item{{\bf CFRS03.0645}}

Also this galaxy shows rotation that is orthogonal to the photometric
major axis, which may indicate an outflow and the $\sigma$-map has a
peak at the edge of the galaxy. We classified the kinematics as
complex. Zheng et al. (2005) found a relatively blue color all over
the galaxy and classified it as a probable merger.\\

\item{{\bf CFRS22.0919}}

Its kinematics is classified as complex: the velocity field is perturbed
and the $\sigma$-map does not show any peak. Note the tails, characteristic 
of interacting systems, seen in the
HST image.\\

\item{{\bf CFRS22.0975}}

Its kinematics is classified as complex: the velocity field is perturbed
and the $\sigma$-map shows a peak but not at the galaxy center. HST
imaging reveals 3 distinct components. This system is probably just
preceding a merger and the high velocity gradient in Figure \ref{Fig1}
can probably not be attributed to rotation.\\

\item{{\bf CFRS03.0523}}

Its kinematics is classified as complex: the velocity field and the
$\sigma$-map do not show any kind of structure expected from rotation.
HST imaging shows a tidal tail, probably indicating ongoing
interactions or gas accretion. This galaxy has a central region bluer
than the outer region (Zheng et al. 2004). The $\sigma$ maxima on the
outer edges of the galaxy correspond to relatively blue regions (see
Zheng et al. 2004).\\

\item{{\bf HDFS4130}}

Its kinematics is classified as complex: the velocity field shows rotation
but the $\sigma$-map does not have any peak. This galaxy is asymmetric
with (spiral ?) arms visible on one side only (maybe a distant
version of a ``tadpole'' galaxy).\\

\item{{\bf HDFS4090}}

Its kinematics is classified as complex: the velocity field is perturbed
and the $\sigma$-map does not have any peak. The morphology looks like a
``peanut''.\\

\item{{\bf HDFS5140}}

Its kinematics is classified as complex: the velocity field is very
perturbed, although the $\sigma$-map has a peak near the center.\\

\item{{\bf HDFS5030}}

Its kinematics is classified as complex: both the velocity field and the
$\sigma$-map appear perturbed.\\
\end{description}

In the sample, 53\% of the LCGs (9 galaxies) present complex
kinematics, which are very different from classical nearby spiral
galaxies (e.g. Garrido et al. 2002; 2003 \& 2004) or early-type
galaxies (Emsellem 2004), indicating that these LCGs are unrelaxed
systems. As Figure \ref{Fig1} shows, there is no correlation between
these dynamical classes and their SNR. Even discarding the lowest SNR
pixels would not change the nature of the kinematics and hence the
kinematical complexity of a galaxy cannot be attributed to a lower
SNR.

Five galaxies, or 29\%, have perturbed kinematics and the remaining
three galaxies (18\%) look like normal spiral galaxies. Interestingly,
among these last eight galaxies (perturbed rotation and rotating
disks), all but three (CFRS03.0619, HDFS5150 and HDFS5190) have
possible nearby companions which might indicate interactions
responsible for their compactness (see Barton \& van Zee 2001).

\begin{figure}[h!]
\centering
\includegraphics[width=9cm]{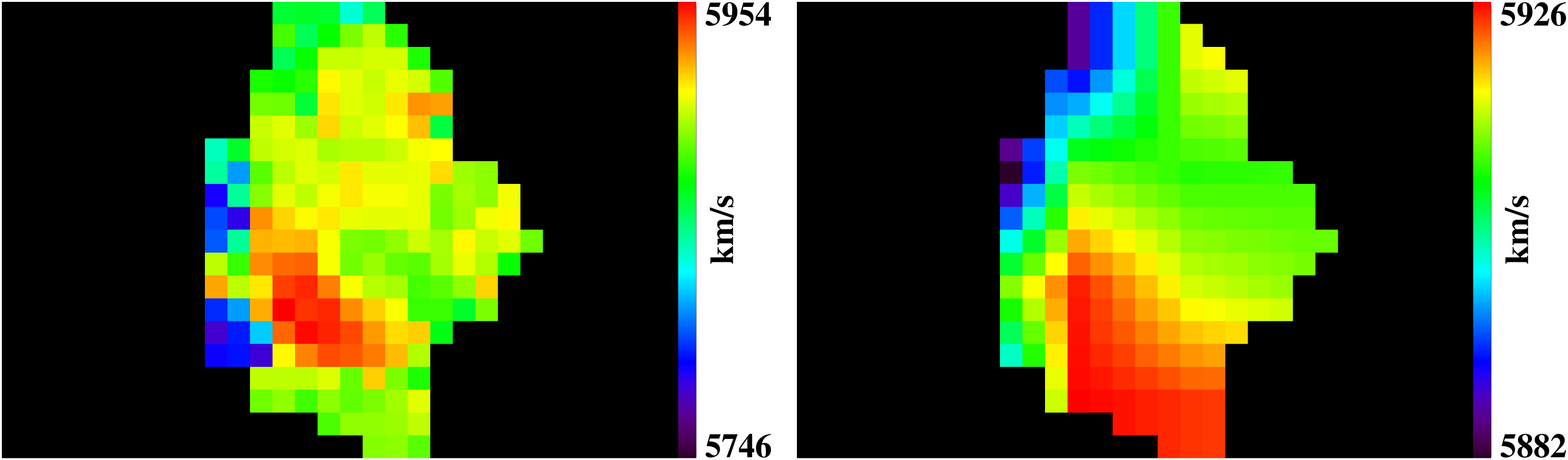}
\caption{Simulation of the velocity field of a LCG at z=0.6. Left:
Redshifted velocity field of ESO400-G43 (see \"Ostlin et al. 1999)
with a $\sim$ 0.1 arcsec pixel (1/5 of a GIRAFFE pixel). Right:
velocity field of the same galaxy as could have been seen by GIRAFFE
(0.6 seeing and 0.52 arcsec/pixel) with a simple 5x5 interpolation and
assuming the same spatial extension. The resolution is dramatically
reduced as well as the velocity gradient (by a factor 4.7), but no
artificial features are created.}
\label{Fig2}
\end{figure}

Before concluding on the dynamical nature of LCGs, we have to
investigate whether such perturbed/complex velocity fields could be
artificial features caused by the low spatial sampling of GIRAFFE. In
Flores et al. (2004 \& 2006), we illustrated the ability of GIRAFFE to
recover regular velocity fields of distant rotating disks. The
question is then to demonstrate the ability of GIRAFFE to recover also
more complex velocity fields in distant galaxies. To tackle this
issue, we used Perot-Fabry observations by \"Ostlin et al. (1999) of a
local LCG. We redshifted the ESO\,400-G43 data cube (see \"Ostlin et
al. 1999 \& 2001) to z=0.6 and simulated the effects of a 0.52 arcsec
sampling and a 0.6 arcsec seeing. The result is illustrated in Figure
\ref{Fig2}: GIRAFFE smoothes the velocity fields (and thereby
underestimates the velocity gradient) but no artificial features are
created. Moreover, the simulated velocity field shows some
similarities with those of the LCGs in Fig. 1.

We can then conclude that LCGs are mainly (53 to 82\% as judged
from our rather small number statistics) unrelaxed systems, at least
when ionized gas motions are considered.
%\"Ostlin et al.
%(2004) showed that in one local LCG gas and stars seem dynamically
%decoupled, and that stars seem more relaxed that gas. 
Such kind of complex kinematics have also been observed by Swinbank et
al. (2005) who obtained the velocity field of a merger remnant at $z
\sim$ 0.1, similar to the complex velocity fields of our LCGs.
%However, the complex kinematics of most LCGs is not sufficient to
%establish firmly the nature of LCGs, i.e. disentangling between a
%merger or a dE scenario (see Introduction): if LCGs are progenitors of
%dE undergoing very strong star formation episodes, it is not clear
%whether their kinematics should appear relaxed.
However, this conclusion relies on the kinematics of the gas
\emph{only}: \"Ostlin et al. (2004) showed that in one local LCG, gas
and stars seem dynamically decoupled, and that stars seem more relaxed
than the gas. In the following section, we attempt to investigate the
nature of the LCG kinematics, namely if they are mostly dominated by
rotation or by dispersion. The fact that LCGs are not dynamically
relaxed systems is a severe limitation for this exercise.

%__________________________________________________________________ 
\section{Test of the dynamics of LCGs}

To investigate energy balances of LCGs, we assume in this section that
  LCGs are systems almost at equilibrium. In this case, one could
  imagine two possibilities: (1) LCGs are mainly supported by velocity
  dispersion, or (2) LCGs are mainly supported by rotation. Both
  options require to assume that LCGs are dynamically not too far from
  equilibrium and we would like to point out again that this
  assumption represent a severe limit in setting up energy balances of
  LCGs. Moreover, the last possibility (that LCGs are dominated by
  rotation) requires in addition to identify large scale motions to
  rotation, which is even more speculative and will thus be detailed
  in appendix.

\subsection{Can LCGs be supported by velocity dispersion?}
If LCGs were systems at equilibrium and supported by velocity
dispersion, then the intensity-weighted mean $\sigma$ and the one
derived from their integrated spectra $\sigma _{intg}$ should be in
agreement (e.g. Bershady et al. 2004; \"Ostlin et al. 2001).

We constructed integrated spectra by the direct summation of all the
spatial elements in the whole IFU. Integrated [OII] lines were fitted
following the same way as for IFUs pixels, but due to the influence of
larger scale motions (rotation for spirals) that widen integrated
lines, about $50\%$ of the integrated spectra were impossible to fit
correctly by a double Gaussian. We then summed independently the fits
of each line of the [OII] doublet over the whole IFU, fitted both by
single Gaussian (corrected from instrumental dispersion) and estimated
the integrated velocity dispersion taking the mean of the two velocity
dispersions derived independently. We checked that in most cases,
these two velocity dispersions were very similar and that when
integrated line fit was possible, both methods gave similar results.

\begin{figure}[h!]
\centering \includegraphics[width=9cm]{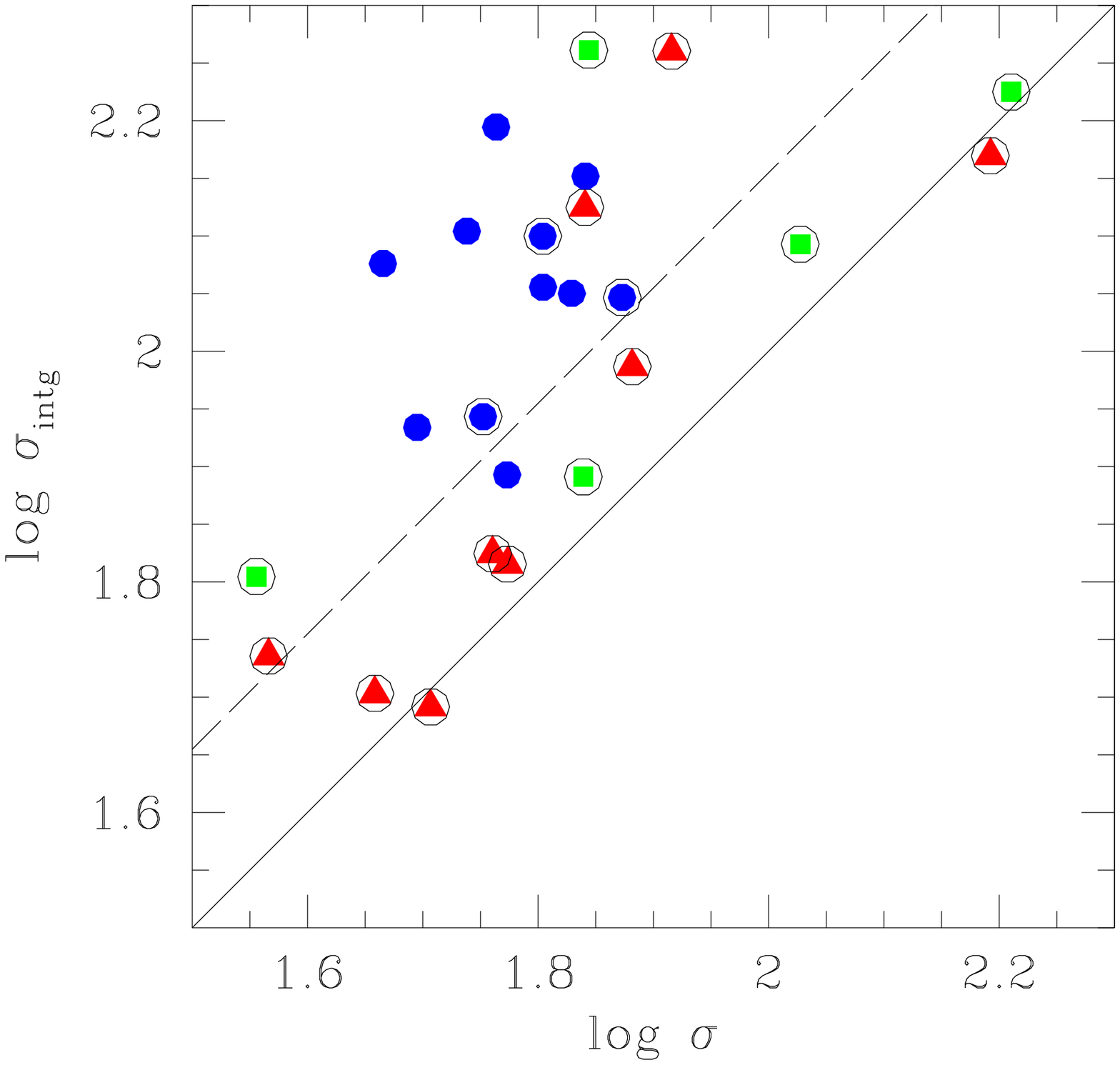}
\caption{Velocity dispersion $\sigma_{intg}$ derived from integrated
spectra vs intensity weighted mean velocity dispersion $\sigma$
derived from $\sigma$-maps. Encircled blue full dots are LCGs
classified as rotating disks, encircled green squares are LCGs
classified as perturbed and encircled red triangles are complex LCGs.
Blue full dots are rotating disks from Paper I (Flores et al. 2006),
added for comparison. The dash line represents the limit where the
relative difference between $\sigma$ and $\sigma _{intg}$ is 30\%.}
\label{Fig9}
\end{figure}

Both measurements are presented in Table \ref{tab2} and Figure
\ref{Fig9}. We estimated the error on the sigma measurement to be 10\%
(median, see Paper I). We thus adopt a 30\% relative treshold (3-sigma
treshold) between $\sigma$ and $\sigma _{intg}$ to identify the
galaxies which could potentialy be supported by velocity dispersion.
Nine galaxies (roughly 50\%) have such a relative difference between
$\sigma$ and $\sigma _{intg}$: CFRS03.0523, CFRS03.1032, CFRS22.0919,
HDFS4090, HDFS5030, HDFS5150, CFRS03.0645 CFRS22.1064 and HDFS4130
(see Table \ref{tab2} and Figure \ref{Fig9}). Note that among these
galaxies, CFRS03.1032 is a very peculiar case because of both its very
high central $\sigma$ and its AGN activity (see individual comments).

%If these are indeed galaxies in
%equilibrium supported by velocity dispersion, then some structure
%hould be observed in their kinematics, at least in their
%$\sigma$-map. However, this is not the case (see Figure \ref{Fig1})
%and we classified all these galaxies as complex: such kinematics could
%hardly be the one of galaxies at equilibrium and supported by velocity
%dispersion as stated by Bershady et al. (2004).

%Figure \ref{Fig9} illustrates the fact that a majority of LCGs cannot
%be supported by velocity dispersion, because their integrated spectra
%demonstrate that their energy balances are dominated by large scale
%motions.

{\scriptsize
\begin{table*}
\centering
\begin{tabular}{llrrrr}\hline\hline
ID    & class & $V_{rot}$ &  $\sigma$  & $\sigma _{intg}$ & $M_*$\\\hline
HDFS4170      & RD   & 173 & 75  & 111 & 10.83 \\
HDFS5190      & RD   & 168 & 64  & 126 & 10.51  \\
CFRS03.0619   & RD   & 155 & 57  & 88  & 10.50 \\\hline
CFRS03.1032   & PR*  & 139 & 162 & 168 & 10.87 \\
CFRS22.0619   & PR   & 67  & 36  & 64  & 9.50  \\
CFRS03.1349   & PR*  & 235 & 70  & 183 & 10.91 \\
CFRS22.1064   & PR   & 124 & 107 & 124 & 10.39 \\
HDFS5150      & PR   & 77  & 69  & 78  & 10.14 \\\hline
CFRS03.0508   & CK   & 78  & 37  & 54  & 9.87  \\
CFRS03.0645   & CK   & 125 & 58  & 67  & 10.27 \\
CFRS22.0919   & CK   & 45  & 46  & 51  & 9.54  \\
CFRS22.0975   & CK   & 409 & 82  & 182 & 10.82 \\
CFRS03.0523   & CK   & 97  & 156 & 148 & 10.35 \\
HDFS4130      & CK   & 153 & 76  & 97  & 10.63 \\
HDFS4090      & CK   & 20  & 51  & 49  & 9.67  \\
HDFS5140      & CK   & 224 & 69  & 133 & 9.91  \\
HDFS5030      & CK   & 74  & 59  & 65  & 10.81 \\\hline\hline
\end{tabular}
\caption{Dynamical properties of the sample of LCGs. The column
entries are (from left to right): id, dynamical class (RD=Rotating
Disk, PR=Perturbed Rotation, CK=Complex Kinematics), maximal
rotational velocity (corrected from inclination but not from GIRAFFE
spatial sampling; The objects for which the kinematical classification
is uncertain are indicated by a star, see individual comments),
intensity-weighted velocity dispersion calculated from $\sigma$-map,
velocity dispersion derived from integrated spectra and stellar masses
derived from photometry (in solar mass).}
\label{tab2}
\end{table*}
}

\subsection{Dynamical support of LCGs}
The other possibily for LCGs to be systems at equilibrium is to be
supported by rotation. This would imply that the large scale motions
seen in the velocity fields of LCGs are due to rotation, althgough
these motions are not completely relaxed. However, we already know
that this is not true for at least the objects whose dynamical axis
are not aligned with the optical one, ie for objects suspected of
outflows (see individual comments). In the appendix, we nevertheless
naively assumed that the large scale motions are precisely due to
rotation. As we assumed that LCGs are at equilibirum, we can then set
up energy balance using classical relations linking their kinematics
(velocity or dispersion) to their mass, to see if these balances are
effectively dominated by the large scale motions interprated as
rotation. Under such an assumption, one finds that 70\% of LCGs have
an energy balance effectively consistent with rotation. However, this
question can only be properly addressed with the knowledge of the
kinematics of the stars (see e.g. \"Ostlin et al. 2004).

To summarize, we find that about 50\% of LCGs could be supported by
velocity dispersion, assuming they are not too far from equilibrium.
Given that 18\% of LCGs are classified as RD, it remains one third of
LCGs for which we cannot exclude that a rotational support could play
a role in their dynamical state. Both large scale and random motions
seem to play an important role in about 40\% of LCGs. Such a mix is
perfectly compatible with mergers.

%Among these 38\%,
%17\% might have an non-negligeable contribution from dispersion. Such
%a mix between large scale motions and random motions is perfectly
%compatible with mergers.

%__________________________________________________________________ 
\section{Stellar vs. dynamical masses}
\begin{figure}[h!]
\centering
   \includegraphics[width=9cm]{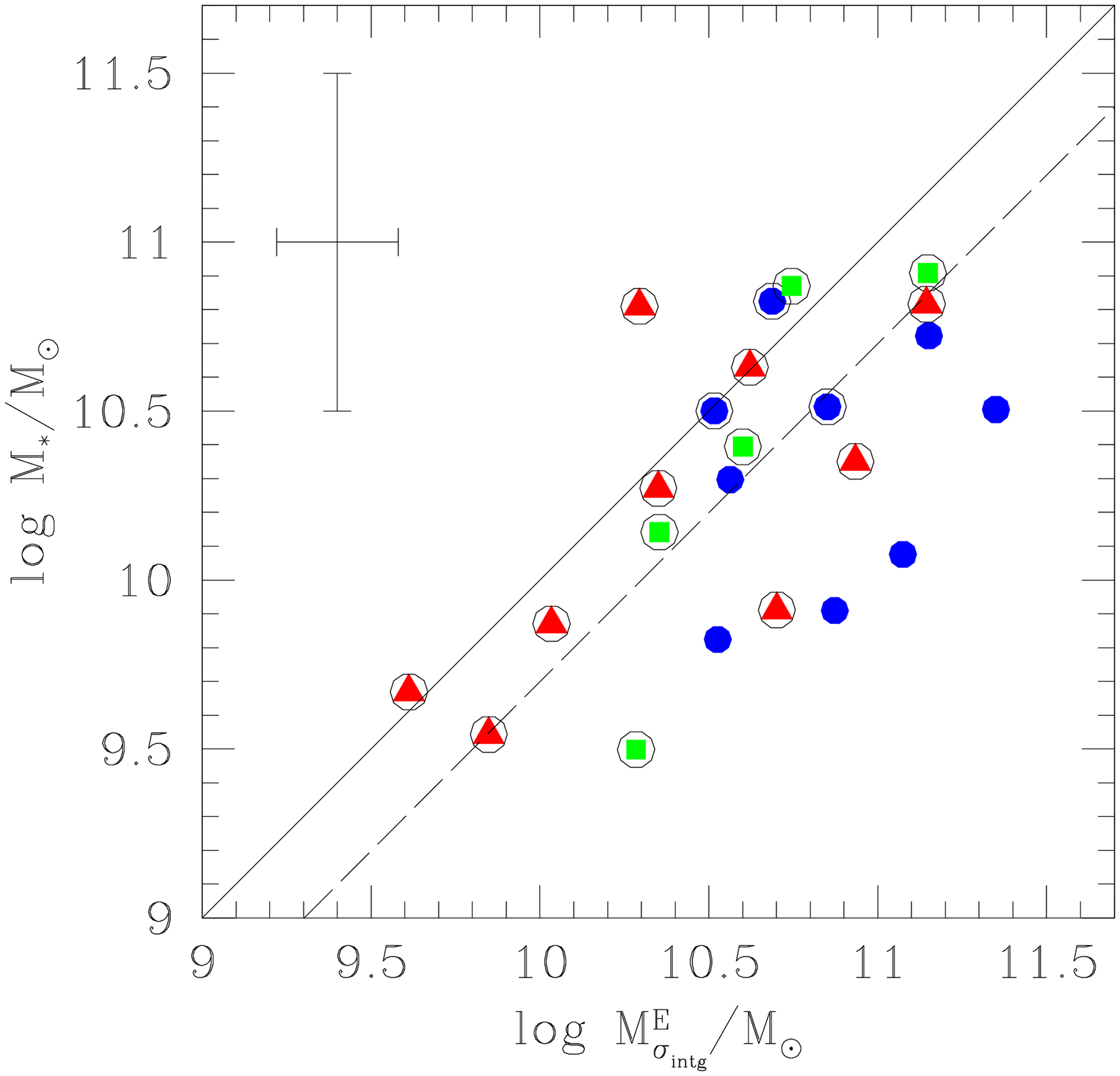}
      \caption{Stellar masses vs dynamical ${\cal M}^E$. Encircled
      blue full dots are LCGs classified as rotating disks, encircled
      green squares are LCGs classified as perturbed and encircled red
      triangles are complex LCGs. Blue full dots are rotating disks
      from Paper I (Flores et al. 2006), added for comparison. For
      convenience, only the median error bar has been plotted.}
         \label{Fig6}
   \end{figure}

We derive stellar masses $M_*$ following Hammer et al. (2005) (see
Table 2). Figure \ref{Fig6} compares stellar masses with total
dynamical pseudo-equivalent masses derived from integrated spectra,
which is the most reliable estimate of the total dynamical mass we can
use (see Appendix). Following Conselice et al. (2005) who carried an
extensive study of uncertainties and systematics on stellar masses, we
estimate our error on stellar masses to be 0.5 dex. Among the sample
of spirals of Paper I (Flores et al. 2006), a mean ratio of dynamical
to stellar masses of $\sim$ 6 (median $\sim$ 7) was found, which is
roughly in agreement with the value of Conselice et al. (2005) who
found a mean ratio of $\sim$ 5 for a sample of $z \sim$0.5 disk
galaxies. The difference between these two samples of spirals is most
probably related to the difference of observational strategy (slit
spectroscopy vs Integral Field Spectroscopy), because the maximal
rotational velocity can be largely underestimated when measured by
slit spectroscopy, since only a part of the whole kinematics is
sampled (see Discussion). On the other hand, we find for our sample of
LCGs a mean ratio between dynamical ${\cal M}^E$ and stellar masses of
$\sim$ 2 (median $\sim$ 1.6). If the majority LCGs are merging
systems, as suggested by their kinematics, their spectra are then
likely dominated by unrelaxed motions, which could lead to
underestimate their total dynamical ${\cal M}^E_{\sigma_{intg}}$.

%__________________________________________________________________
\section{Discussion \& conclusion}
We have presented GIRAFFE-IFU observations and derived velocity fields
 and $\sigma$ maps for 17 LCGs. Our main result is that 18\% have
 velocity fields characteristic of rotating disks kinematics, the rest
 having perturbed or complex kinematics. This result is unlikely to be
 affected if we account for the four objects that were discarded
 because of a too low SNR. Among them, two show (in HST images) tails
 characteristics of ongoing interactions. Assuming that they are not
 too far from equilibrium, we find that about half of LCGs could be
 supported by velocity dispersion. The remaining 32\% of LCGs seems to
 be dominated by large scale motions. An important fraction of LCGs,
 between 32 and 82\%, are thus probable mergers.

In this sample of 17 LCGs, four have their velocity gradient nearly
perpendicular to their main optical axis (CFRS03.0508, CFRS03.0645,
CFRS22.0975 and HDFS5140). It suggests that in these galaxies, gas is
tracing feedback processes such as outflows, rather than gravitational
dynamics: in this case, the dynamical axis should appear off axis
relative to the optical main axis (see Veilleux et al. 2005 and
references therein). All of them except CFRS22.0975 have stellar
masses lower than $3.10^{10}M_\odot$ which is the upper threshold for
supernova feedback to drive efficient outflows (Dekel \& Birnboim
2005). In CFRS22.0975, the gas velocity field probably traces the
relative motion of the merger progenitors (see its morphology in
Figure 1). For CFRS03.0645, we retrieved FORS slit spectra (Hammer et
al. 2001) to compare systemic velocities of emission ($H_\beta$,
$H_\gamma$, [OII] and [OIII]) and absorption lines (CaII, H and K) and
found no significant shifts between them, which makes the outflow
hypothesis uncertain for this galaxy. Note however that the slit has
been aligned with the apparent (optical) main axis which could explain
why we did not see any shift between emission and absorption lines.
The electron density map of CFRS03.0508 is presented in Paper III
(Puech et al. 2006b) and support the outflow hypothesis.

%{\bf As suggested by \"Ostlin et al. (2004), it might be that stars in
%LCGs could be more relaxed than gas, and that LCGs could then be
%systems not too far from equilibrium. We thus tried to push our
%analysis further by naively assuming that LCGs are at equilibrium. We
%then find that only 30\% can be supported by dispersion because of
%evidence of large scale motions in the integrated spectra of the
%remaining 70\%. If one identify these large scale motions with
%rotation, it is found that at least 70\% of LCGs have energy balances
%that are be dominated by rotation. Given that we are likely
%underestimating the rotational velocity of kinematical complex LCGs,
%we cannot exclude that {\em all} LCGs could have an energy balance
%dominated by rotation. This conclusion entirely rely on the assumption
%that large scale motions are due to rotation. Several indirect clues,
%such as the comparison between dynamical ${\cal M}^E$ and ${\cal M}^E$ derived from
%integrated spectra or the comparison of the stellar masses to
%dynamical ${\cal M}^E$ between rotating disks and LCGs, suggest that this
%could be the case. However, this possible link between LCGs and
%rotating disks remains very uncertain.}

Bershady et al. (2004) used the STIS long slit spectrograph
  onboard the HST to study the kinematics of 6 LCGs with \(M_B \sim
  -21\), \(r_{half}\sim2\) \(h_{70}^{-1}\) kpc and $\sigma \sim$65
  km/s, and found that LCGs are supported by velocity dispersion.
  Unfortunately, only one galaxy of the present sample (CFRS220919) is
  enough compact to fulfill these selection criteria. Interestingly,
  it has being pointed out by Hammer et al. (2001) that this galaxy
  could be a possible progenitor of a dE. Here we find that this
  galaxy is a possible candidate for being supported by dispersion.
  However, given the complexity of most LCGs kinematics presented
  here, slit spectroscopy should be used with care for this kind of
  objects. A careful inspection of Figure \ref{Fig1} reveals that with
  long slit spectroscopy, the real nature of many LCGs presented in
  this paper would have been misinterpreted. In the case of
  CFRS03.0645, a slit positioned along the main optical axis would
  have revealed a rather flat velocity gradient and would have
  completely missed the rotation. The case of CFRS03.0508 is even more
  instructive: the same exercise would have revealed a flat velocity
  gradient and a clear $\sigma$ peak, and this galaxy would then have
  been classified as supported by dispersion.
%Given the small number
%  of LCGs in their sample, we believe that such a bias could become
%  dramatic and may explain why they did not see any rotation in their
%  LCG sample.

%These results substantially differ from those of Bershady et al.
%(2004) who found that LCGs are supported by velocity dispersion.
%However, they used the STIS long slit spectrograph on-board the HST and
%given the complexity of most LCGs kinematics, one could easily
%understand their results because they did not sample the whole
%kinematics. A careful inspection of Figure \ref{Fig1} reveals that
%with long slit spectroscopy, the real nature of many LCGs presented in
%this paper would have been misinterpreted. In the case of CFRS03.0645,
%a slit positioned along the main optical axis would have revealed a
%rather flat velocity gradient and would have completely missed the
%rotation. The case of CFRS03.0508 is even more instructive: the same
%exercise would have revealed a flat velocity gradient and a clear
%$\sigma$ peak, and this galaxy would then have been classified as
%supported by dispersion. Given the small number of LCGs in their
%sample, we believe that such a bias could become dramatic and may
%explain why they did not see any rotation in their LCG sample.

Finally, how can we interpret the compactness of LCGs? Most of
perturbed kinematics and rotating LCGs show possible companions with
which they could be in interaction. This could explain their
compactness as due to interactions and/or minor mergers, following
Barton \& van Zee (2001, see introduction). Another conjecture is the
one of Hammer et al. (2005). They proposed a scenario where local
massive spiral could form after major mergers in three main phases.
The sequence would start by a pre-merging phase during which the
system would form a huge amount of stars and appear as a LIRG. The
second phase would be the LCG phase, were all material falls onto the
mass barycenter of the merging system which could enhanced the star
formation activity in the center of these systems, making them looking
compact. This is consistent with Bergvall \& \"Ostlin (2002) who found
central intense starbursts superimposed on low surface brightness
components in four local BCGs. Our results are consistent with this
picture as we found that most LCGs are objects with complex kinematics
as expected from major mergers. During the third phase of the
scenario, a disk would grow thanks to material accreted from the IGM.
Inflows/outflows are also predicted by this scenario, arising from
feedback and gas falling back to form a new disk (see Robertson et al.
2005).
%On the contrary, our
%results are clearly incompatible with the dE picture proposed by Koo
%et al (1995), Guzman et al. (1997) and Bershady et al. (2004) since we
%find no evidence that LCGs could be supported by dispersion. 
%Our results thus suggest that LCGs are interacting or merging systems.

How could we distinguish between minor and major mergers? In a minor
merger, the disk cannot be destroyed and the kinematics of the remnant
cannot appear too complex. We should then observe a galaxy still
rotating along its main optical axis but a dispersion map peaking
outside the center, where the smaller progenitor falls. This could
correspond to the LCGs we classified perturbed. On the other hand,
during a major merger the disk is completely destroyed or at least
strongly perturbed. In such a case, we should then observe either a
rotation significantly misaligned with the optical axis combined with
a non-centered dispersion map, or a complex kinematics without any
obvious structure. We emphasize that given our spectral resolution,
reaching $\sim$ 10000, the fact that the [OII] doublet is not always
resolved reveals by itself the complexity of some of these galaxies.

LCGs dominate the evolution of the UV luminosity from $z=$1 to 0.
Their role can thus not be negligible in the process of formation and
evolution of galaxies during the last 8 Gyr. Our results highly
support a hierarchical type picture where galaxies form from smaller
units. In this picture, LCGs seem to be a major event as proposed by
Hammer et al. (2005). Although based on small numbers, our sample is
nevertheless representative of the galaxy population at \(0.4\leq
z\leq 0.75\) (see Paper I). A larger sample is under construction as
part of the ESO Large Program IMAGES (P.I.: F. Hammer) and will be a
decisive step towards the confirmation of the spiral rebuilding
scenario. Recently, it has been shown that major merger remnants may
not necessarily be ellipticals but also spirals, depending on the gas
abundance (Springel and Hernquist 2005) and/or the Black Hole feedback
(Robertson et al. 2005). Comparisons with theoretical simulations will
bring a crucial test of the spiral rebuilding scenario and on the
nature of LCGs.

%__________________________________________________________________
\begin{acknowledgements}
We thank P. Amram and C. Balkowski for their help and very useful
comments, and our R. Guzman, our referee, for his very useful comments
and suggestions. We also thank A. Bosma for his enlightening comment
on CFRS03.0508. We are especially indebted to T.J. Cox who provide us
with an hydrodynamical simulations of a Sbc galaxy. HF and MP wish to
thank ESO Paranal staff for their reception and their very useful
advises during observations. We thank all the team of GIRAFFE at Paris
Observatory, at Geneve Observatory and at ESO for the remarkable
accomplishment of this unique instrument, without which, none of these
results would be obtained.
\end{acknowledgements}

%__________________________________________________________________

\appendix

\section{Energy balance of LCGs}

This appendix is devoted to the energy balance of LCGs. It is
explicitly assumed that LCGs are systems at equilibrium and supported
by rotation. We will thus assume in this appendix that the large scale
motions seen in the velocity fields are associated with rotation, even
if the true origin of these large scale motions is largely uncertain,
except for some suspected outflows and those clearly identified as RD
from their kinematics (see individual comments and section 4.2). We
will nevertheless naively assume that their origin is rotation and see
if any contradiction arises. To test this hypothesis, we set up an
energy balance in the sample of LCGs. In this balance, we will take
into account the contributions from large scale ordered motions
(interpreted as rotation) and from random motions. Energies are
estimated in what we call ``pseudo equivalent masses'' (${\cal M}^E$, i.e. in
mass units). It is important to emphasize that these ${\cal M}^E$ can be
interpreted as real masses {\it only} for galaxies in equilibrium.

%Given the results of the last section (that
%LCGs cannot be systems in equilibrium supported by velocity
%dispersion) and their kinematics (see Figure \ref{Fig1}), we test if
%LCGs could be described by hidden, more or less perturbed, rotating
%disks. We then interpret the large scale motions as due to rotation
%and use classical relations for spiral galaxies to derive rotational
%nd velocity dispersion masses.}

\subsection{Contribution from rotation}
\begin{figure}[h!]
\centering
\includegraphics[width=9cm]{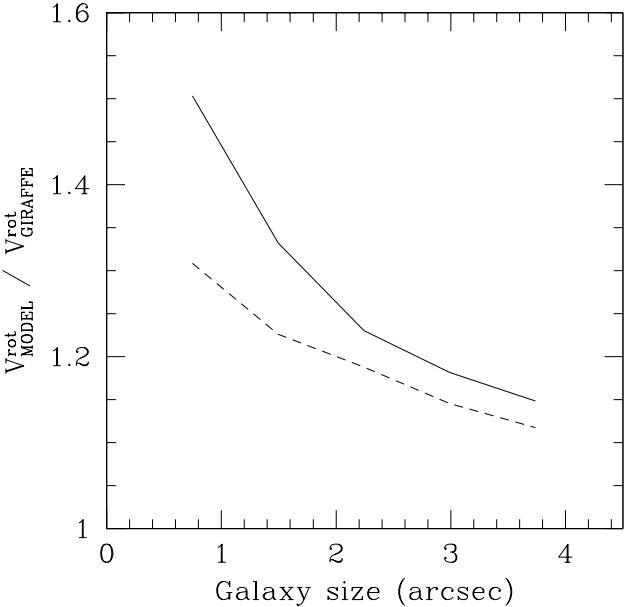}
\caption{Correcting factor for the maximal rotational velocity vs
galaxy size. Due to its low spatial sampling, the GIRAFFE IFU leads
to underestimate $V_{rot}$ by a factor $1.2\pm 0.04$ for galaxies
sizes between 2 and 3 arcsec. Full line: tilted view ($\sim 53$
degrees). Dash line: top view ($\sim 25$ degrees)}
\label{Fig3}
\end{figure}

To estimate the mass supported by rotation, we assumed that the
maximal rotational velocity is equal to half the maximal gradient of
the velocity field, corrected for inclination (see Table 1). Due to
the low spatial sampling and large distances to our targets, we have
already pointed out that the GIRAFFE/IFU observations will
underestimate the maximal rotational velocity (see Figure \ref{Fig2}).
Using hydrodynamical simulations of an Sbc (Milky-Way like) galaxy by
Cox et al. (2004), we simulated our GIRAFFE observations assuming
median atmospherical conditions at ESO VLT (0.81 arcsec seeing at 500
nm and an outer scale for the turbulence of 24 m, see Tokovinin 2002).
We scaled this template galaxy ($V_{rot}=160$ km/s and $inc=53$
degrees) to fill boxes of length ranging from 0.75 to 6 arcsec to
mimic distant galaxies, and then compared the kinematics seen by
GIRAFFE with the original simulation (see Figure \ref{Fig3}). We found
that for spiral galaxies with sizes between 2 and 3 arcsec, GIRAFFE is
able to correctly recover the maximal rotational velocity, although it
is underestimated by $\sim$ 20 \%. For more complex kinematics, the
correction factor should be larger (see also Figure \ref{Fig2}):
between 2 and 6. However, as we explicitly assume in this section
that LCGs are rotating disks, we applied a constant factor of 20\%
whatever the dynamical class (spiral or perturbed/complex) of LCGs in
order to compute homogeneous estimates.

Rotational ${\cal M}^E$ are then estimated from (Lequeux 1983):
$${\cal M}^E_{v}=f{R_{25}V_{rot}^2 \over G}$$ 

with f=0.6 for a disk with a flat rotation curve and
$R_{25}=2r_{half}$, following Phillips et al. (1997). For a pure
exponential disk with scale length $r_d$, $r_{half}=1.67 r_d$. Taking
$R_{25}=2r_{half}$ in fact assumes that $r_{half}/r_d= 1.6$ since
$R_{25}=3.2r_d$ (Persic \& Salucci 1988). With ${\cal M}^E$ in ${\cal
M}_\odot$, $V_{rot}$ in km/s and $r_{half}$ in kpc, this becomes:
$${\cal M}^E_{v}=0.279\cdot 10^6r_{half}V_{rot}^2$$

Corrected rotational velocities and ${\cal M}^E$ are given in Table 2.
To estimate our error bars we took into account the uncertainties on
$r_{half}$ ($\sim$1/2 HST pixel which represents 0.34 kpc at z=0.6),
$V_{rot}$ ($\pm$ 5 km/s, determined by repeating several times the
fitting procedure), inclination ($\pm 4$ degrees, see section 2) and
the correction factor on the velocity ($\pm 0.04$, see Figure
\ref{Fig3}). Note however, that the real uncertainty on inclination is
probably higher for unrelaxed systems, as its derivation usually
relies on the assumption of a thin disk seen in projection (see
section 2). From this we estimated an median error on ${\cal M}^E_{v}$
of $\sim$ 0.16 dex which is mainly dominated by the uncertainty in the
inclination.

{\scriptsize
\begin{table*}
\centering
\begin{tabular}{llrrrrr}\hline\hline
ID    & class & $\sigma _{corr}$ & ${\cal M}^E_v$ & ${\cal M}^E_\sigma$ & ${\cal M}^E_{dyn}$ & ${\cal M}^E \sigma _{intg}$ \\\hline
HDFS4170      & RD   & 56  & 10.63 & 10.10 &  10.74 &  10.69 \\
HDFS5190      & RD   & 27  & 10.66 & 9.50  &  10.69 &  10.85 \\
CFRS03.0619   & RD   & 55  & 10.57 & 10.10 &  10.70 &  10.52 \\\hline
CFRS03.1032   & PR*  & 161 & 10.14 & 10.71 &  10.81 &  10.75 \\
CFRS22.0619   & PR   & 24  & 9.89  & 9.43  &  10.02 &  10.29  \\
CFRS03.1349   & PR*  & 34  & 10.93 & 9.70  &  10.94 &  11.15 \\
CFRS22.1064   & PR   & 101 & 10.17 & 10.43 &  10.62 &  10.60  \\
HDFS5150      & PR   & 67  & 9.90  & 10.23 &  10.40 &  10.35  \\\hline
CFRS03.0508   & CK   & 34  & 9.91  & 9.63  &  10.09 &  10.03  \\
CFRS03.0645   & CK   & 49  & 10.46 & 10.09 &  10.60 &  10.35  \\
CFRS22.0919   & CK   & 43  & 9.30  & 9.70  &  9.85  &  9.85   \\
CFRS22.0975   & CK   & 10  & 11.41 & 8.62  &  11.41 &  11.15  \\
CFRS03.0523   & CK   & 155 & 10.13 & 10.97 &  11.03 &  10.93  \\
HDFS4130      & CK   & 72  & 10.58 & 10.36 &  10.79 &  10.62  \\
HDFS4090      & CK   & 51  & 8.38  & 9.64  &  9.66  &  9.61   \\
HDFS5140      & CK   & 32  & 10.71 & 9.45  &  10.74 &  10.70  \\
HDFS5030      & CK   & 59  & 9.96  & 10.21 &  10.40 &  10.29  \\\hline\hline
\end{tabular}
\caption{Dynamical properties of the sample of LCGs. The column
entries are (from left to right): id, dynamical class (RD=Rotating
Disk, PR=Perturbed Rotation, CK=Complex Kinematics),
intensity-weighted velocity dispersion corrected from GIRAFFE sampling
effect (see text), ${\cal M}^E$ supported by rotation (corrected from
GIRAFFE sampling effect), ${\cal M}^E$ supported by random motions,
total dynamical ${\cal M}^E$, and dynamical ${\cal M}^E$ estimated by
integrated velocity dispersion }
\label{tab3}
\end{table*}
}

\subsection{Contribution from velocity dispersion}
\begin{figure}[h!]
\centering
\includegraphics[width=9cm]{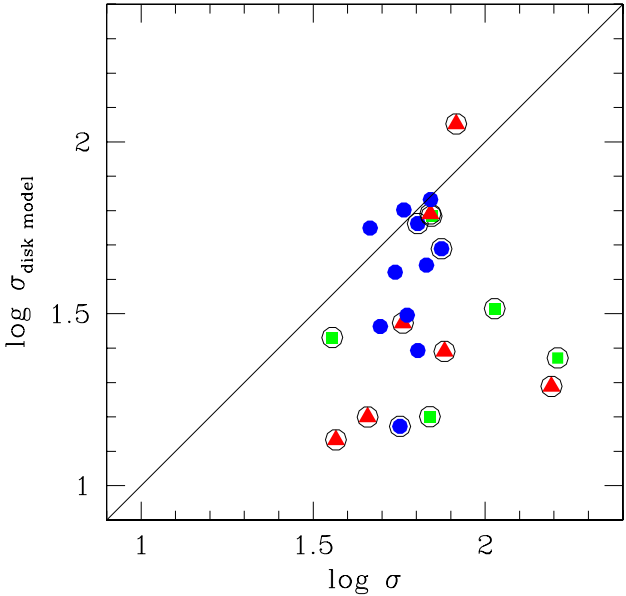}
\caption{Correction term $\sigma_{disk model}$ vs intensity weighted
mean $\sigma$. Encircled blue full dots are LCGs classified as
rotating disks, encircled green squares are LCGs classified as
perturbed and encircled red triangles are complex LCGs. Blue full dots
are rotating disks from Paper I (Flores et al. 2006), added for
comparison.}
\label{Fig7}
\end{figure}

Following \"Ostlin et al. (2001), we estimated the ${\cal M}^E$ supported by
velocity dispersion using (${\cal M}^E$ in ${\cal M}_\odot$, $\sigma$ in km/s
and $r_{half}$ in kpc):
$${\cal M}^E_{\sigma}=1.1\cdot 10^6r_{half}\sigma^2$$ where $\sigma$ is the
intensity weighted mean of the $\sigma$ -map. Due to the low GIRAFFE
spatial sampling, $\sigma$ will tend to be overestimated, since the
coarse pixel size integrates large scale motions (rotation for a
spiral galaxy). We used the same hydrodynamical simulation to estimate
the increase in $\sigma$ due to large scale motions and found $\sim$
50 km/s ($V_{rot}=160$ km/s and $inc=53$ degrees). We used another
simulation to check that the effect approximatively scales with
$1/\sin(inc)$ (within an error on $\sigma$ of 5 km/s) and assumed the
same scaling with $V_{rot}$. We then corrected the GIRAFFE intensity
weighted mean $\sigma$ using this recipe and from this estimated the
mass supported by velocity dispersion (see Table \ref{tab3} and Figure
\ref{Fig7}). For pure rotating disks, a correction roughly equal to
the measure is expected, which means that intrinsic velocity
dispersion is negligible. Galaxies classified as rotating disk and
which fall far from the $\sigma = \sigma _{disk model}$ region could
be galaxies with a significant bulge (e.g. CFRS03.9003, see also Puech
et al. 2006b) or show perturbation in their $\sigma$-map (CFRS03.0619,
CFRS22.0504 and HDFS4020). Interestingly, these 4 galaxies (plus
CDFS03.1032, but see individual comments) are among those which are
the nearest of the $\sigma _{intg}= \sigma$ line on Figure \ref{Fig9},
which supports the idea that their dynamical support could include a
substantial contribution coming from dispersion. All points with a
correction larger than the measured mean $\sigma$ are spiral galaxies
except one (CFRS22.0975) whose velocity gradient is likely due to
relative motions between merging components (see its morphology in
Figure \ref{Fig1}) rather than rotation. In these few cases, we fixed
the corrected value to 10 km/s which corresponds to the minimal
expected line width due to intrinsic turbulent motions in spiral
galaxies (Rozas et al. 1998; Van Zee \& Bryant 1999). We estimate a
median error on ${\cal M}^E_\sigma$ of $\sim$ 0.08 dex.

\subsection{Total dynamical ${\cal M}^E$}
To estimate total dynamical pseudo-equivalent masses ${\cal M}^E_{dyn}$, we
simply added ${\cal M}^E_v$ and ${\cal M}^E_\sigma$. Most authors derive $\sigma$ on
integrated spectra (global velocity dispersion) and use the relation
of the previous section to estimate the whole dynamical mass
${\cal M}^E_{\sigma _{intg}}$ (Guzman et al. 1996, Phillips et al. 1997).
Under assumptions concerning the anisotropy of the kinetic energy
tensor and the geometry of the system, this relation can be used to
estimate total dynamical masses of rotating and flattened spheroids
(Bender et al. 1992). This approach is often used in studies using
slit spectroscopy (Guzman et al. 1997; Guzman et al. 2001; Hammer et
al. 2001) but the validity of this relation for systems dominated by
rotation is uncertain. In the following, we also estimated
${\cal M}^E_{\sigma _{intg}}$ which we compared with ${\cal M}^E_{dyn}$.

\begin{figure}[h!]
\centering \includegraphics[width=9cm]{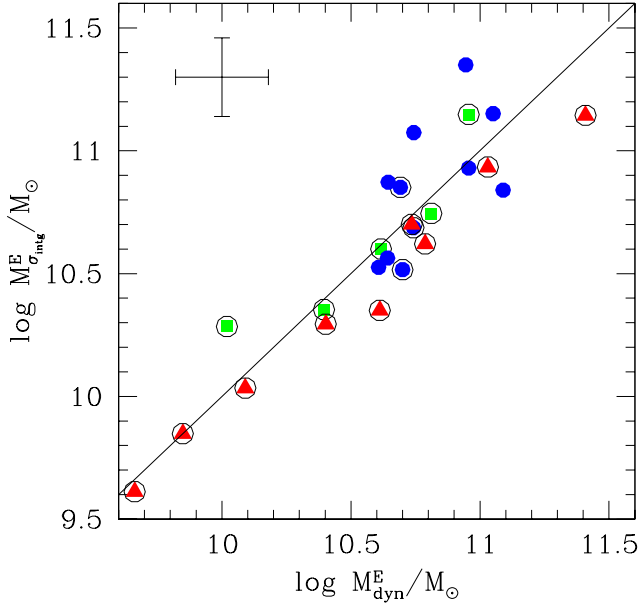}
      \caption{Dynamical ${\cal M}^E$ derived from integrated spectra
      vs dynamical ${\cal M}^E$ derived from 3D spectroscopy.
      Encircled blue full dots are LCGs classified as rotating disks,
      encircled green squares are LCGs classified as perturbed and
      encircled red triangles are complex LCGs. Blue full dots are
      rotating disks from Paper I (Flores et al. 2006), added for
      comparison. For convenience, only the median error bar has been
      plotted.}
         \label{Fig5}
   \end{figure}

Figure \ref{Fig5} shows the comparison between ${\cal M}^E_{dyn}$ and
${\cal M}^E_{\sigma _{intg}}$. We find a correlation between the two
estimates, which seem to validate our estimates and is consistent with
the fact that LCGs could be systems not too far from equilibrium.
Spiral galaxies are almost equally distributed on both sides of the
line where ${\cal M}^E_{\sigma _{intg}}={\cal M}^E_{dyn}$ which is
likely due to the fact that ${\cal M}^E_{\sigma _{intg}}$ does not
correct explicitly for inclination effects.
% We choose to use our
%estimate of ${\cal M}_{dyn}$ in the following.

\begin{figure}[h!]
\centering
   \includegraphics[width=9cm]{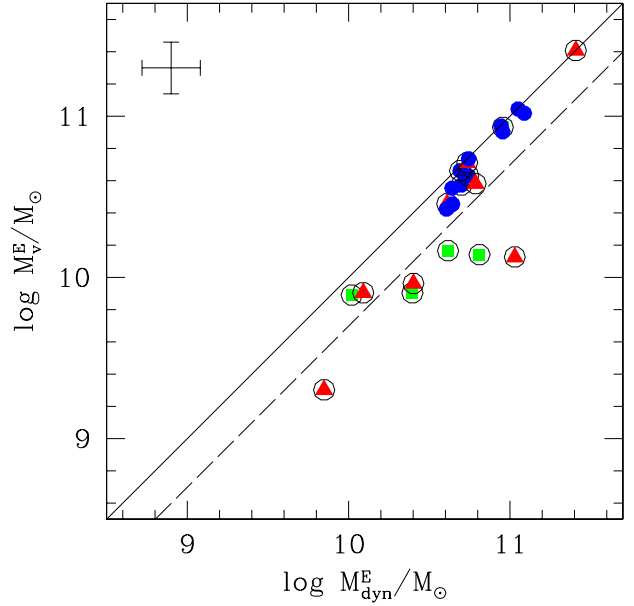}
      \caption{Rotational ${\cal M}^E$ vs total dynamical ${\cal
      M}^E$. Encircled symbols are the sample of LCGs. Encircled blue
      full dots are LCGs classified as rotating disks, encircled green
      squares are LCGs classified as perturbed and encircled red
      triangles are complex LCGs. Blue full dots are rotating disks
      from Paper I (Flores et al. 2006), added for comparison. The
      dash line represent the limit where ${\cal M}^E_v$=${\cal
      M}^E_{dyn}$/2. For convenience, only the median error bar has
      been plotted. Notice that the high ${\cal M}^E_v$ of CFRS22.0975
      (the point at the top-right) could not be attributed to a normal
      rotation since it is an obvious merger.}
         \label{Fig4}
   \end{figure}

In Figure \ref{Fig4} we compare ${\cal M}^E_v$ with ${\cal M}^E_{dyn}$
to test if the LCGs are dominated by rotation. For comparison, we also
plot a sample of 8 rotating disks taken from Paper I (Flores et al.
2006). Clearly, the whole spiral sample falls in the rotation
dominated area. At least 70\% of the LCGs (12 galaxies) seems to be
dominated by rotation whereas the remaining 30\% (5 galaxies) seems to
be dominated by velocity dispersion. Note that these five galaxies
were already identified as potentially supported by dispersion (see
section 4.1): CFRS22.0919, CFRS03.0523, CFRS03.1032 HDF4090 and
CFRS22.1064. Interestingly, one of these galaxies (CFRS22.0919) was
identified by Hammer et al. (2001) as a potential dwarf progenitor as
discussed by Guzman et al. (1997).
%We
%have already pointed out that because these galaxies have complex
%kinematics, they cannot be equilibrium systems supported by
%dispersion, so these objects cannot follow the interpretation of
%Bershady et al. (2004). 
Because we are likely underestimating their rotational velocities, we
cannot exclude that all LCGs have energy balance dominated by
rotation.

However, we recall that this conclusion relies on the assumptions that
large scale motions in the velocity fields are due to rotation, and
that we know that for some systems (e.g. the suspected outflows) this
is far from being true. We nevertheless cannot exclude that a
possible relation could link most of the LCGs with rotating disks.

\end{document}